\documentclass[10pt,twocolumn]{article} 
\usepackage{simpleConference}
\usepackage{times}

\pdfoutput=1

\usepackage{graphicx}
\usepackage{amssymb}
\usepackage{url}

\usepackage[hyperfootnotes=false]{hyperref}

\usepackage{float} 

\usepackage{multirow}
\usepackage{siunitx}

\usepackage{booktabs,caption}
\usepackage[flushleft]{threeparttable}

\usepackage{bm}        
\usepackage{amssymb}   
\usepackage{amsmath}   
\makeatletter
\newcommand{\mathleft}{\@fleqntrue\@mathmargin20pt}
\newcommand{\mathcenter}{\@fleqnfalse}
\makeatother

\begin{document}

\vspace{2cm}

\title{On hadron deformation: a model independent extraction of EMR from  pion photoproduction data}

\author{L. Markou$^{1}$, E. Stiliaris$^{2}$ and C. N.Papanicolas$^{1,*}$ \\
\\
$^{1}$ The Cyprus Institute, K. Kavafi 20, 2121 Nicosia, Cyprus \\ 
$^{2}$ National and Kapodistrian University of Athens, Physics Department, 15771 Athens, Greece \\
\today
\\
\\
}


\twocolumn[
  \begin{@twocolumnfalse}
    \maketitle
    \begin{abstract}
    The multipole content of pion photoproduction at the  $\Delta^+ (1232)$ resonance has 
been extracted from a data set dominated by recent Mainz Microtron (MAMI) precision measurements. 
The analysis has been carried out in the Athens Model Independent Analysis Scheme (AMIAS), 
thus eliminating any model bias. 
The benchmark quantity for nucleon deformation, $EMR = E2/M1 = E_{1+}^{3/2}/M_{1+}^{3/2}$, was determined to 
be $-2.5 \pm 0.4_{stat+syst}$, thus reconfirming in a model independent way 
that the conjecture of baryon deformation is valid. 
The derived multipole amplitudes  provide stringent constraints on 
QCD simulations and QCD inspired models striving to describe hadronic structure. 
They are in good agreement with phenomenological models which explicitly incorporate pionic degrees of freedom 
 and with lattice QCD calculations.
\end{abstract}

\textbf{PACS.} 13.60.Rj -Baryon production – 14.20.Gk -Baryon resonances $(S=0)$ – 24.10.Lx Monte Carlo simulations – 25.20.Lj Photoproduction reactions 
\\
\\

  \end{@twocolumnfalse}
]

\thispagestyle{empty}


\section{Introduction}
\label{intro}

The conjectured deformation of the nucleon \cite{glashow1979unmellisonant, papanicolas2007shapes,RevModPhys.84.1231} and of hadrons in general,  
has been the focus of numerous theoretical and experimental investigations for over thirty years 
\cite{PhysRevLett.78.606,blanpied1997n,frolov1999electroproduction,beck2000determination,blanpied2001n,PhysRevLett.86.2963,bartsch2002measurement,PhysRevLett.88.122001,ahrens2004helicity,sparveris2005investigation,Beck2006,kotulla2007real,dreschsel1992threshold,PhysRevC.63.055201,kamalov2001pi,aznauryan2003multipole,alexandrou2011nucleon} as it addresses a fundamental question in physics: the shape of the smallest particles in nature known to have size. 

The anticipated deformation of hadrons is attributed to fundamental QCD processes: the color hyperfine interaction and its tensor component in particular\cite{isgur1982n,PhysRevD.41.2767,ramalho2008d, pascalutsa2007electromagnetic}  gives rise to hadron deformation in a similar manner that nucleon-nucleon tensor interaction gives rise to the deuteron deformation. It is also recognized that equally fundamental complex chiral dynamics are at work which lead to the deformation of the nucleon's pion cloud. These same fundamental features of QCD are also responsible for many other aspects of hadron structure such as the non-vanishing neutron charge RMS radius and the magnitude of their magnetic moments.
 
The resulting non spherical components, predominantly D-wave admixtures, in the nucleon wavefunction allow quadrupole excitation of the $\Delta$ which would be absent if only S-waves were present.
The ratio of electric quadrupole $E2$ to magnetic dipole $M1$ amplitudes (EMR) measured in the $\gamma p \to \Delta^+(1232)$ transition serves as the accepted gauge of the magnitude of the deformation of the proton \cite{papanicolas2007shapes}.  QCD simulations in lattice gauge theories and QCD inspired models yield non vanishing EMR. However both the theoretical calculations and the experimental results need to reach higher accuracy and precision to guide theoretical efforts for a better understanding of this fundamental issue. 

 Invariably, the  experimental investigations to measure EMR utilize the transitions: 
 \begin{equation}
  \label{eq:transition}
  \gamma p \to \Delta^+(1232) \to \begin{cases}
                n\pi^{+}  \\
                p \pi^{0} \\
                \gamma p
            \end{cases}
\end{equation}
where a photon ($\gamma$), real or virtual,  excites a proton ($p$)  to a $\Delta^{+} (1232)$ which 
 de-excites with the emission of a pion ($\pi$) or a gamma ray. The $\Delta^{+}(1232)$
resonance  $(J=I=3/2, \Gamma = 117MeV)$ \cite{Patrignani:2016xqp} 
 decays 99.4$\%$ to the  $\pi N$ channel and
$(0.55-0.65)\%$ to the $\gamma$ channel \cite{Patrignani:2016xqp}. Its shape and width are dominated by the $\pi N$ interaction.

As the final state of the $\gamma N \to \pi N$  lies in the continuum,  the number of possible contributing multipoles 
is very large, in principle infinite.
Apart from the resonant multipoles, all other multipoles are termed as ``background''  deriving primarily 
from Born terms and from the tails of higher resonances \cite{dreschsel1992threshold}. 
Due to insufficient data \cite{wunderlich2016complete,W11} and limitations of the analysis 
techniques \cite{AMIAS_benchmark}  multipole analyses up to now have been able to
extract only few multipoles. Higher multipoles are either set to zero (truncated analysis) or  
use models to account for them (model dependent analysis). This model dependence, unavoidably, 
leads to shifted mean values and underestimated uncertainties in the extracted multipoles and the EMR. 
It has been argued that in the $\Delta(1232)$ region the model error could significantly 
influence the results \cite{bernstein2007overview,arndt2001multipole} making it difficult 
to extract multipoles with the necessary accuracy so as to provide constraints and guidance to the various theoretical models or to address the issue of nucleon deformation.

In the work presented here both limitations have been overcome: 
a) A rich and precise dataset has been assembled which includes the most recent pion photoproduction data 
to date \cite{adlarson2015measurement,schumann2015threshold,otte,annand2016t}
and adequate experimental observables to allow a model independent multipole extraction and 
b) the Athens Model Independent Analysis Scheme (AMIAS) \cite{stiliaris2007multipole,papanicolas2012novel} has been 
employed for 
the analysis of the data. For the first time AMIAS inherent capability to incorporate systematic error in 
the analysis and treat it on the same footing as statistical error has been implemented. 
Thus, the results presented in this work derive from the most recent and most accurate photoproduction data available, 
some of them used for the first time for multipole analysis  \cite{adlarson2015measurement,schumann2015threshold,otte,annand2016t}, 
employing the most complete analysis scheme currently available.

In this paper  multipoles are classified using the standard $\pi N$ notation in which
a multipole is noted as $X_{\ell_{\pi} \pm}^{I}$ where $X=E, M$   indicates its character - electric or magnetic, $J$ its total angular momentum,  $I$ its isospin,
$\ell_{\pi}$  its  orbital angular momentum; ``$+$'' or ``$-$'' is used to denote whether the spin (s=1/2) is parallel or anti-parallel to the angular momentum.

\section{The experimental database}
\label{sec:data}
High accuracy neutral pion photoproduction  data in the $\Delta(1232)$ region
were acquired at MAMI \cite{adlarson2015measurement,schumann2015threshold,otte,annand2016t} by the A2 collaboration \cite{A2}. These data 
include the unpolarized differential cross section $d\sigma_0$ \cite{adlarson2015measurement}, 
 and the polarization-dependent differential cross sections $\sigma T$ \cite{schumann2015threshold,otte,annand2016t} and $\sigma F$ 
 \cite{schumann2015threshold,otte,annand2016t} associated with the target asymmetry $T$ and the beam-target asymmetry $F$ respectively. 
The $\hat{T}=\sigma T$ observable was measured with a transversely polarized  target \cite{schumann2015threshold,otte,annand2016t} 
and it features higher statistical precision and angular coverage over earlier measurements \cite{dutz1996photoproduction}. 
 The  $\hat{F}=\sigma F$  observable was measured for the first time using a 
  transversely polarized target with a longitudinally polarized beam \cite{schumann2015threshold,otte,annand2016t}.

  The  $d\sigma_0$ \cite{adlarson2015measurement} is the most precise $\gamma p \to p \pi^{0}$ cross section measured to 
date characterized by unprecedented statistical accuracy and 
extended angular coverage. It features a fine binning in $E_{\gamma}$ of $4MeV$ and covers the full pion production angle in 30 angular bins. 
The polarization data feature $1.1MeV$ bins in $E_{\gamma}$ and $18$ 
evenly spaced angular bins from $5^{\circ}$ to $175^{\circ}$. 
All aforementioned data  were measured with the use of the Glasgow-Mainz photon tagging 
facility \cite{anthony1991designGlasgow,hall1996focalGlasgow,mcgeorge2008upgradeGlasgow}. 
The Crystal Ball \cite{starostin2001measurementCrystalBall} and the TAPS multiphoton detector \cite{gabler1994responseTAPS,novotny1991bafTAPS} 
served as central and forward calorimeters respectively. 

\begin{figure}[htp]
\centering  
{\includegraphics[width = 3.2in]{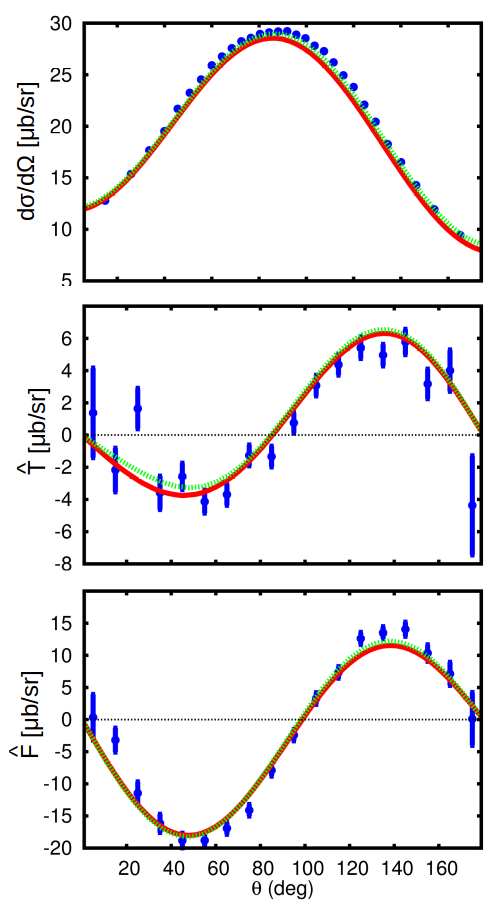}} \hfil 
\vspace*{-3mm}
\caption{The high precision MAMI data used in our analysis at $W_{cm}=1232MeV$. From top to bottom are 
the  $\gamma p \to p\pi^0$ differential 
cross section \cite{adlarson2015measurement}, the polarized target $\hat{T}$ observable \cite{schumann2015threshold,otte,annand2016t} and  
the first ever measurement of the polarized beam-target $\hat{F}$ observable \cite{schumann2015threshold,otte,annand2016t}. The red solid  
and the green dashed curves are the  MAID07 prediction and the SAID (CM12) solution at the same energy.}
\label{fig:paperepjdata}
\end{figure}

Fig. \ref{fig:paperepjdata} shows the unpolarized  $d\sigma_0$  and 
the $\hat{T}$ and $\hat{F}$ polarization data; the experimental observations are in good agreement with both the 
MAID07 model (red solid curve) and the SAID (CM12) solution 
(green dashed curve).  Only statistical errors are shown.

In our analysis 
the recent MAMI data were complemented by the older but still most precise $\gamma p \to n \pi^{+} $ 
data of Beck \textit{et al.} \cite{beck2000determination} for 
unpolarized cross section ($d \sigma_0$) and beam asymmetry ($\Sigma$) and the double 
polarization beam-target  $G$ and $P$ asymmetries of Ahrens \textit{et al.} \cite{Ahrens2005} 
and Belyaev \textit{et al.} \cite{belyaev1983experimental}. The $G$ asymmetry was measured using a  $4\pi$
detector system,  a linearly polarized tagged photon beam, and a longitudinally polarized proton target. 
$P$ was measured using linearly polarized photons and a transversely polarized proton target \cite{belyaev1983experimental}.
The  $G$ and $P$ measurements feature only a limited number of  
angular measurements of low statistical precision but as it will be shown in section \ref{subsec:bands} they are 
important in restricting  the derived multipole solutions. The analyzed dataset is listed in Table \ref{tab:paperepjdata}. 

\begin{table}[!htpb]
\centering
\caption{The experimental data used in this work. Data  in the $\Delta(1232)$ region presented and used for the first time  are indicated 
by a $^{*}$.} 
  \begin{tabular}{cccc}
    \toprule
        Observable & $E_{\gamma} (MeV)$ & Ref. & datapoints   \\ 
          \hline
    \multicolumn{4}{c}{$\gamma p \to p \pi^{0}$}    \\
  \hline
  $d \sigma_0$ & 337.6 - 342.0  & \cite{adlarson2015measurement}  & 30     \\
  $\Sigma$     & 335 - 345      & \cite{Leukel} & 17          \\
  $T^{*}$     & 339.0 - 340.1       & \cite{schumann2015threshold,otte,annand2016t} & 18           \\
  $F^{*}$     & 339.0 - 340.1       & \cite{schumann2015threshold,otte,annand2016t} & 18    \\
  $G$     & 326 - 354           & \cite{Ahrens2005}  & 3           \\
  $P$     & 335 - 365           & \cite{belyaev1983experimental} & 6       \\    
    \hline
  \multicolumn{4}{c}{$\gamma p \to n \pi^{+}$} \\
    \hline
      $d \sigma_0$ & 335 - 345   & \cite{beck2000determination}& 10 \\
       $\Sigma$ & 335 - 345               &  \cite{beck2000determination}& 10 \\
        $T$ & 335 - 356       &  \cite{dutz1996photoproduction}  & 11 \\
        $G$ & 335 - 356                &   \cite{Ahrens2005}  & 6 \\
         $P$ & 330 - 350     & \cite{get1981positive} & 6  \\
  \bottomrule
  \end{tabular}
  \label{tab:paperepjdata}
\end{table}

\section{Methodology}
\label{sec:meth}
The methodology employed is the implementation of the  Chew, Goldberger, Low and Nambu (CGLN) 
theory \cite{chew1957relativistic} in the Athens Model Independent Analysis Scheme (AMIAS) \cite{stiliaris2007multipole,papanicolas2012novel} for 
use with  photoproduction data. In the case of  single pion photoproduction, 
$\gamma N \to N\pi$, both initial particles and the final state nucleon have two spin states 
yielding a total of eight degrees of freedom \cite{beck2000determination,dreschsel1992threshold}. 
Due to parity conservation a total of four complex amplitudes are required to describe the reaction \cite{beck2000determination}. 
The four invariant amplitudes of the photoproduction process are related to the CGLN amplitudes  $\left ( F_1, F_2, F_3, F_4 \right ) $ 
by a linear and invertible transformation \cite{chew1957relativistic,HANSTEIN1998561}.

The AMIAS method is based on statistical concepts and relies heavily on Monte Carlo and simulation techniques, 
and it thus requires High Performance Computing as it is
computationally intensive. The method identifies and determines with maximal 
precision parameters that are sensitive to the data by yielding their Probability Distribution Function (PDF). 
The AMIAS is computationally robust and numerically stable. It has been successfully applied in the analysis of 
data from nucleon electroproduction resonance \cite{stiliaris2007multipole,RevModPhys.84.1231}, 
lattice QCD simulations \cite{alexandrou2015novel} and medical imaging 
\cite{loizospseudo}. 

AMIAS requires that the parameters 
to be extracted from the experimental data are explicitly linked via a theory or a model \cite{stiliaris2007multipole}. 
This  requirement is fulfilled, like  in the case of electroproduction ref. \cite{stiliaris2007multipole,markou}, 
as  multipoles are connected to the pion photoproduction observables via the CGLN \cite{chew1957relativistic} amplitudes.
The multipole series of the CGLN amplitudes $\left ( F_i, i=1,4 \right ) $ takes the form \cite{chew1957relativistic}:
\mathleft
\begin{equation}  \label{eq:cgln1}
 \begin{split}
 F_{1}  = {} &  \sum_{l=0}^{\infty}   [ \left ( lM_{l+}  + E_{l+} \right ) P'_{l+1}(x)   \\
  & + \left ( \left (l+1 \right ) M_{l-} + E_{l-} \right ) P'_{l-1}(x) ]
\end{split}
\end{equation}
\begin{equation}
  \label{eq:cgln2}
 F_{2}  = {} \sum_{l=1}^{\infty} \left [\left (l+1 \right)  M_{l+}  + lM_{l-}  \right ]  P'_{l}(x)
\end{equation}
\begin{equation}   \label{eq:cgln3}
 \begin{split}
  F_{3}  = {} & \sum_{l=1}^{\infty}  [ \left ( E_{l+} - M_{l+}  \right ) P''_{l}(x) \\ 
  & + \left ( E_{l-}  + M_{l-}  \right ) P''_{l-1}(x)  ]   
\end{split}
\end{equation}
\begin{equation}
 F_{4}  = {} \sum_{l=2}^{\infty} \left [ M_{l+}  - E_{l+}  - M_{l-}  - E_{l-} \right ] P''_{l-1}(x)  
 \label{eq:cgln4}
\end{equation}
 where $x = \cos(\theta)$ is the cosine of the scattering angle. 
It is also assumed that in addition to unitarity the use of the 
Fermi-Watson theorem \cite{FW} applies below the two-pion threshold which in turn implies that the multipole phases in photoproduction are 
equal to the  $\pi N$ scattering phase shifts \cite{workman2012parameterization}. By fixing the multipole phases from the 
experimentally determined $\pi N$ phases \cite{stiliaris2007multipole,workman2012parameterization}
the parameters of the problem become definite isospin multipole amplitudes, namely, the $A^{1/2}$ and $A^{3/2}$ 
amplitudes  \cite{dreschsel1992threshold}. These amplitudes are obtained from the reaction channel amplitudes and the relations \cite{dreschsel1992threshold}:
\mathcenter
\begin{align}
A^{1/2} &={} \frac{A_{p \pi^{0}}}{3} +  \frac{\sqrt{2} A_{n \pi^{+}}}{3}, & A^{3/2}  &={}A_{p \pi^{0}} - \frac{A_{n \pi^{+}}} {\sqrt{2}}
\label{eq:epjaiso}
\end{align}

\begin{figure}[htp]
\centering  
{\includegraphics[width = 3.4in]{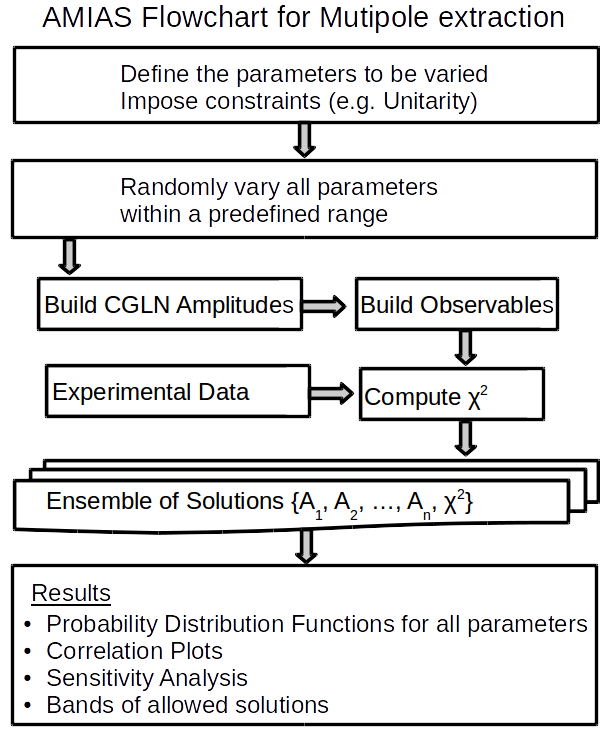}} \\
\caption{AMIAS flowchart for multipole extraction. The term parameters  includes multipoles, nuisance parameters and kinematical variables.}
\label{fig:amias_flow}
\end{figure}

The Fermi-Watson theorem  provides a particularly useful constraint enabling the model independent analysis 
as it has been shown that the number of discrete ambiguities in unconstrained 
truncated multipole analyses rises exponentially with the number of multipoles being fitted \cite{omelaenko1981ambiguities,wunderlich2014complete}. 
In contrast to the practice adhered up to now 
where multipoles which are not fitted  are either fixed 
through a model \cite{sparveris2005investigation,davidson2001model} or through their Born contribution \cite{beck2000determination}, we exploit 
the AMIAS's robustness and  numerical stability to extract all multipole amplitudes to which the data exhibit any sensitivity.
This feature of AMIAS is discussed in \cite{alexandrou2015novel}  and it is demonstrated in this work  
 by extracting multipole amplitudes by  gradually increasing the $\ell_{cut}$, where $\ell_{cut}$ is the upper summation limit 
of eq. $(2)$ - $(5)$.  
All multipoles of order higher than $\ell_{cut}$  which are not varied are fixed to their Born contribution which is calculated 
up to all orders \cite{priveTiator}. We adhere to the MAID convection \cite{dreschsel1992threshold} and definitions 
concerning "Born terms", which thus 
contain  s- and u-channel nucleon terms, $\rho$ and $\omega$ exchange 
in the t-channel, and pion pole contributions \cite{dreschsel1992threshold} 
which are significant in the charged pion channel. 
Above a given multipole order the derived PDFs remain unchanged as more amplitudes are allowed to vary, indicating that convergence has been reached 
and maximal information has been extracted from the data. As a convergence criterion for the PDFs we use a T-statistics test \cite{barlow1993,porter2008testing} 
and demand that the derived $\ell_{cut}$ multipole 
PDFs differ no more than $\sim 2\%$ from the $\ell_{cut}+1$ PDFs. 
The minimum  $\chi^2$ value generated during  each $\ell_{cut}$ analysis  also reaches an asymptotic value as the analyses are driven 
towards convergence. This means that allowing higher waves to vary would not contribute to the $\chi^2$ value of the problem, as the data are 
completely insensitive to them. A  flowchart of the AMIAS implementation for multipole extraction from photoproduction data is 
illustrated in Fig. \ref{fig:amias_flow}.

Multipole extraction from photoproduction data is an inverse problem posing 
a highly complex \cite{blanpied2001n} and correlated parameter space \cite{AMIAS_benchmark,fernandez2009unexpected}. Capturing 
all these correlations is essential 
 for dealing with the numerous 
 and individually weakly  contributing
 background amplitudes and producing precise results. 
 Model dependent methods freeze ``insensitive''  multipoles thus excluding any possibility 
 of determining them and more importantly removing their influence on the dominant amplitudes, through correlations, 
   which can be substantial. 
 Such an approach introduces uncontrolled  model error  which may both shift the extracted values for 
 the dominant amplitudes and underestimate the corresponding uncertainty. 
 Implementing the AMIAS postulate \cite{stiliaris2007multipole,papanicolas2012novel}, 
 that ``every physically accepted solution is a solution to the problem 
with a finite probability of representing reality'', through  unbiased MC sampling of the entire parameter space, 
all possible correlations between the parameters and up to all orders are captured and embedded 
in the AMIAS ensemble of solutions.  The full accounting of correlations is one of the main features of AMIAS that  minimizes the model error.

\subsection{Treatment of systematic errors}
\label{sec:syst}
Modern accelerators and detection instrumentation have allowed  significant improvements in the   quality 
of pion-photoproduction data, often yielding results characterized by statistical errors smaller than the estimated systematic 
uncertainties \cite{adlarson2015measurement}. This requires a  more complex and sophisticated treatment of systematic effects in 
multipole extraction analyses. In  previous analyses, this was either ignored or accommodated by simply adding 
in quadrature the systematic and statistical uncertainties.  
In the $\Delta(1232)$ region, the dominant $M_{1+}^{3/2}$ amplitude 
is very sensitive to  systematic errors of multiplicative nature while the small resonant  $E_{1+}^{3/2}$ amplitude is in addition 
sensitive 
to  angular precision \cite{beck2000determination}. 

To account for possible sources of systematic uncertainty whose leading effect on the data is either of multiplicative  or 
of additive  nature (pedestal) we have introduced nuisance parameters for the unpolarized cross section data and the  model:  
\mathcenter
\begin{equation}
 d \sigma_{0}^{i} \to \alpha_{i} \cdot d \sigma_0^{i} + c_{i}
\end{equation}
\mathleft
where the coefficients (nuisance parameters) $\alpha_{i}$ and $c_{i}$ are allowed to vary in a restricted range 
according to the magnitude of the reported estimated systematic uncertainty \cite{beck2000determination,adlarson2015measurement}. 
The index-$i$ is used to distinguish between the $p \pi^0$ and the $n \pi^+$ differential cross sections. 
An uncertainty in determining the center-of-mass (CM) pion angle of up to $2^{\circ}$  is reported for the Crystal Ball/Taps system \cite{Collicott}  
 and it was taken into account by allowing the  CM angle that freedom  during the variation (AMIAS Monte Carlo) procedure. 
The uncertainty in incident photon energy $E_{\gamma,lab}$ and center-of-mass energy $W_{cm}$ \cite{adlarson2015measurement,mcgeorge2008upgradeGlasgow}
 was found to induce a  negligible  effect to the data at the resonance region.

\subsection{Validation of the applied methodology}
\label{sec:validation}
The AMIAS methodology for the case of electro-and photoproduction has been extensively studied and validated 
through the analysis of pseudodata \cite{stiliaris2007multipole,papanicolas2012novel,markou}; also
in several other reactions and cases \cite{alexandrou2015novel,loizospseudo}. 

The case of resonance photoproduction presented here was extensively studied; 
an indicative example of  multipole amplitude  extraction  employing  
the  aforementioned methodology using the  pseudodata of \cite{pseudo} is shown here.  The analyzed pseudo-dataset contains  
high precision simulation data for the differential cross section ($d\sigma_0$), the beam ($\Sigma$), the target ($T$) and the beam-target ($F$) asymmetries 
for the $\gamma p \to p \pi^{0}$ and  the $\gamma p \to n \pi^{+}$ reactions. The data were generated by randomizing the  MAID07 \cite{Drechsel:2007if} 
model as input at the CM energy $W=1232.23 MeV$. They  contain eighteen even spaced angular measurements for each observable 
in the dynamical region $\theta_{cm} \in [5^{\circ}:175^{\circ}]$, where $\theta_{cm}$ is the angle between the incoming photon and the 
outgoing pion  in the CM frame \cite{pseudo}. 

\begin{figure} [htp]
{\includegraphics[width = 3.5in]{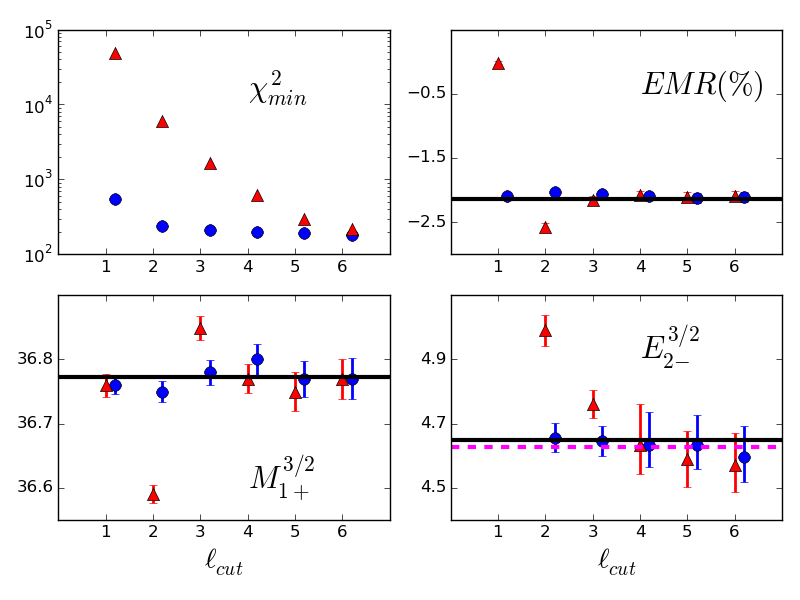}}\hfil 
\vspace*{-6mm}
\caption{From left to right and top to bottom: The minimum $\chi^2$ value generated, 
the extracted $EMR(\%)$, $M_{1+}^{3/2}$ and $E_{2-}^{3/2}$ 
as functions of the angular momentum number $\ell$. 
Red triangles: Truncated analysis. Blue dots: Multipoles which are not varied are fixed by Born terms. The generator values ($MAID07$) 
are marked by a solid black  line ($EMR_{MAID07}=-2.14\%$). The Born contribution to a multipole is marked by a dashed magenta line. }
\label{fig:epjpaperpseudo}
\end{figure}

By applying the methodology of Sec. \ref{sec:meth} we extracted multipole amplitudes of up to $\ell_{cut}=6$ where convergence was reached. 
The $\chi^2_{min}$ found in the AMIAS ensemble  of solutions as the $\ell_{cut}$  increased is illustrated in Fig. \ref{fig:epjpaperpseudo}-a 
where the color-coding  red triangles was used for the truncated analysis (multipoles which are not varied are set to zero) 
while the blue circles were used for the analysis during which amplitudes that were not varied were fixed to their  Born value. 
The derived values for 
$EMR(\%)$ and two selected amplitudes, $M_{1+}^{3/2}$ and $E_{2-}^{3/2}$ are also illustrated from $\ell_{cut}=1$ up to convergence. 
The results shown depict the mean for the derived multipole PDF with an confidence level corresponding to a $\pm 34\%$ probability. 
The black horizontal line is the generator input while the magenta line shows 
the Born  contribution to the multipole. All extracted multipoles  show a similar behavior; 
they are in statistical agreement 
with the generator input and converge by $\ell_{cut}=6$. It is worth noting that the results of the truncated analysis 
 are characterized by large fluctuations as the imposed $\ell_{cut}$ is increased. This indicates that higher waves are needed 
 to reliably extract lower multipoles.

\section{Results}
\label{sec:res}
\begin{figure} [htp]
{\includegraphics[width = 3.4in]{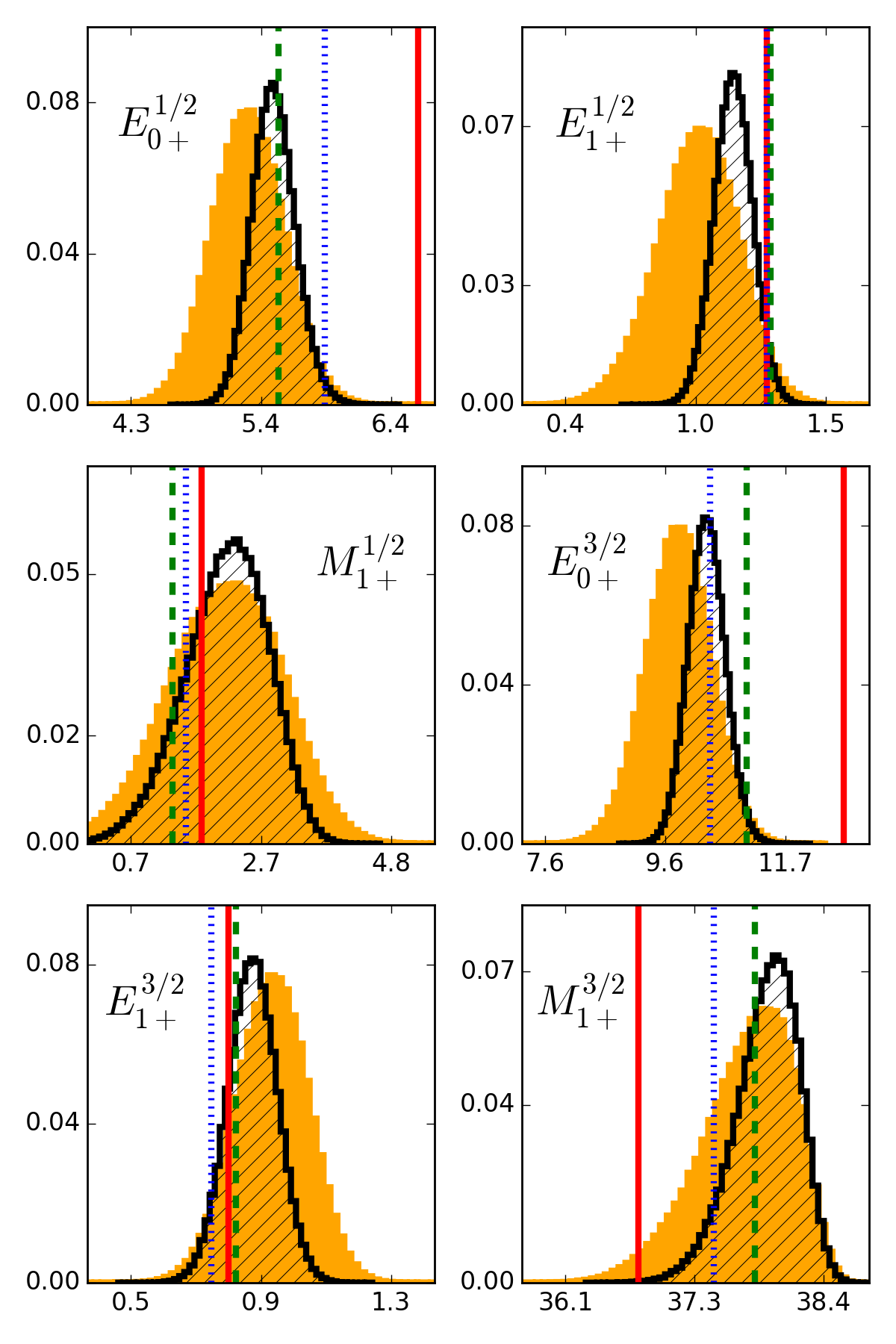}} \\
\vspace*{-5mm}

\caption{PDFs of  $\ell \leq 1$  extracted multipole amplitudes. The AMIAS solution is colored in orange while the 
Model Dependent (MD) $\ell_{cut}=1$  solution in the hatched histogram. 
The  vertical lines show model predictions: MAID07 (red-continuous), SAID-PR15 (green-dashed), and Bonn-Gatchina-2014-02 (blue-dotted).}
\label{fig:spwaves}
 \end{figure}

\begin{figure} [htp]
{\includegraphics[width = 3.4in]{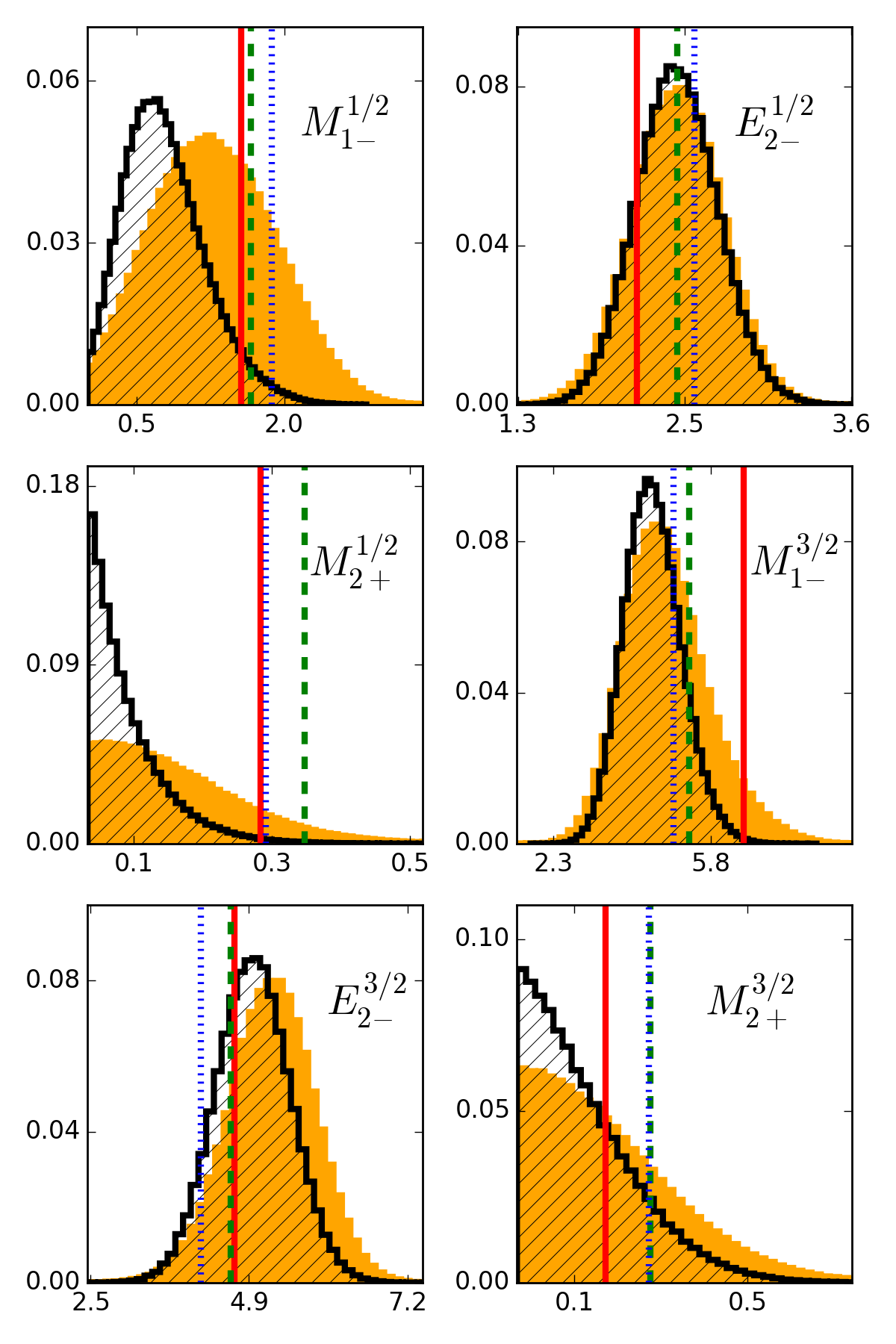}} \\
\vspace*{-5mm}
\caption{PDFs of   $\ell \leq 2$  extracted multipole amplitudes.  Same conventions  as in Fig. \ref{fig:spwaves} are used.}
\label{fig:dwaves}
 \end{figure}

By applying the methodology presented in Section \ref{sec:meth}  
to the experimental data of Table \ref{tab:paperepjdata} we extracted values for all multipole amplitudes which show sensitivity to the data. 
 Multipoles of up to $\ell_{cut}=5$ were varied before convergence was reached. 
In Figs \ref{fig:spwaves} and \ref{fig:dwaves} the PDFs of some selected  $\ell \leq 2$ 
multipoles from the AMIAS model  independent analysis  are compared to the $\ell=1$ and $\ell=2$ model dependent (MD) analysis and 
the Bonn - Gatchina \cite{bgweb,Gutz:2014wit}, the MAID07 \cite{Drechsel:2007if} and the SAID-PR15 \cite{gwuweb2,priveWorkman} solutions. 
 MD analyses recognize the influence of non resonant multipoles of order higher than those allowed by the imposed truncation $\ell_{cut}$, 
 typically  $\ell_{cut}=1$ and $\ell_{cut}=2$, on the derived resonant terms; instead of ignoring them they give 
 them values derived from phenomenological models. In the results shown in Figs \ref{fig:spwaves} and \ref{fig:dwaves} 
 the higher than $\ell_{cut}$ background terms were fixed to the MAID07 values.  
In general, we observe good agreement between the derived mean values between the AMIAS and the MD analyses. We note
 significant differences in the derived uncertainties between the AMIAS and the MD analyses.
 The AMIAS extracted values 
agree with those of the phenomenological models 
which fall within one or two standard deviations from the experimentally 
determined values. 
The sole exception is the MAID07 value for the multipole 
amplitudes $E_{0+}^{1/2}$ and  $E_{0+}^{3/2}$ 
where a larger than $3\sigma$ 
discrepancy is observed. 
The extracted amplitudes with $\ell \geq 3 $  are not 
shown in the figures 
but they are in good statistical agreement 
with  model predictions which treat these background amplitudes 
as primarily   
deriving from Born terms. 

\begin{table}[htp]
\centering
\caption{AMIAS Extracted values and $68\%$ confidence level (CL)  for all 
the multipoles exhibiting sensitivity to 
the data of Table \ref{tab:paperepjdata}. The results are presented
with decreasing accuracy, 
which is the measured  
relative uncertainty (RU). Only multipoles with relative uncertainty 
$<50\%$ are listed.  The MAID07 \cite{Drechsel:2007if}, 
SAID-(PR15) \cite{adlarson2015measurement,priveWorkman} and Bonn-Gatchina \cite{bgweb,Gutz:2014wit} model values are tabulated for comparison.  
In boldface are the predictions which lie outside the $95\%$ CL.
Multipoles are given in units of $10^{-3}/m_{\pi}$.} 
\begin{tabular}{llllcl}
Multipole      & MAID07      &SAID      & BG                     &        AMIAS                         &        RU$(\%)$  \\
\hline
 $M_{1+}^{3/2}$ & {36.7}                &  37.8       		 &  37.3    		&  $    37.7   \pm^{   0.3  }_{   0.5  }  $  &   0.9		\\
 $E_{0+}^{3/2}$ & \textbf{12.7}         &  \textbf{11.0}         &  10.4    		&  $    10.0   \pm^{   0.5  }_{   0.6  }  $  &   5.5			\\  
 $E_{0+}^{1/2}$ & \textbf{6.6}          &  5.5        		 &   \textbf{5.9}       &  $     5.3   \pm^{   0.3  }_{   0.3  }  $  &   5.7		\\  
 $E_{2-}^{3/2}$ & {4.6}      	        &  4.6       		 &   \textbf{4.2}       &  $     5.5   \pm^{   0.6  }_{   0.6  }  $  &  10.9  		\\
 $E_{2-}^{1/2}$ & {2.11}     	 	&  2.6       		 &  2.5    		&  $     2.57  \pm^{   0.27 }_{   0.30 }  $  &  11.1  		\\
 $E_{1+}^{3/2}$ & {0.79}     		&  0.81     		 &  0.73    		&  $     0.94  \pm^{   0.11 }_{   0.13 }  $  &  12.8 	 	\\
 $E_{1+}^{1/2}$ & {1.27}     		&  1.28       		 &  1.26   		&  $     1.07  \pm^{   0.13 }_{   0.16 }  $  &  13.6  		\\
 $M_{1-}^{3/2}$ & {6.6}      		&  5.4      		 &  5.0    		&  $     4.5   \pm^{   1.0  }_{   0.6  }  $  &  17.8		\\
 $M_{2-}^{1/2}$ & {0.55}     		&  0.60     		 &  0.62	        &  $     0.88  \pm^{   0.14 }_{   0.20 }  $  &  19.3 		\\ 
 $M_{1+}^{1/2}$ & {1.8}      		&  1.4      		 &   1.6   		&  $     3.0   \pm^{   0.6  }_{   0.9  }  $  &  25.0		\\
 $E_{2+}^{1/2}$ & {0.36}     		&  0.39     		 &  0.38    		&  $     0.28  \pm^{   0.07 }_{   0.08 }  $  &  26.8  		\\
 $E_{3-}^{1/2}$ & {0.48}     		&  0.46    		 &  0.47   	        &  $     0.42  \pm^{   0.12 }_{   0.12 }  $  &  28.6  		\\
 $E_{4+}^{3/2}$ & {0.08}     		&  \textbf{0.07}         &  \textbf{0.07}       &  $     0.18  \pm^{   0.06 }_{   0.05 }  $  &  30.6  		\\
 $E_{3+}^{3/2}$ & {0.19}     		&  0.19       		 &  0.19   	        &  $     0.28  \pm^{   0.08 }_{   0.10 }  $  &  32.1  		\\
 $E_{3-}^{3/2}$ & {0.61}     		&  0.56      		 &  0.64       	        &  $     0.65  \pm^{   0.25 }_{   0.21 }  $  &  35.4  		\\
 $E_{4-}^{3/2}$ & {0.20}     		&  0.19     		 &  0.19    		&  $     0.40  \pm^{   0.15 }_{   0.14 }  $  &  36.3  		\\
 $E_{3+}^{1/2}$ & {0.13}     		&  0.12       		 &  0.13    		&  $     0.11  \pm^{   0.04 }_{   0.04 }  $  &  36.4  		\\
 $M_{3-}^{3/2}$ & {0.15}     		&  0.14     		 &  0.15    		&  $     0.38  \pm^{   0.16 }_{   0.14 }  $  &  39.5  		\\
 $M_{2-}^{3/2}$ & {0.54}     		&  0.42    		 &  0.59    		&  $     0.63  \pm^{   0.31 }_{   0.23 }  $  &  42.9  		\\
 $E_{2+}^{3/2}$ & {0.55}     		&  0.65     		 & 0.56     		&  $     0.39  \pm^{   0.23 }_{   0.13 }  $  &  46.2 		\\
 $M_{3-}^{1/2}$ & {0.11}     		&  0.12      		 & 0.10     		&  $     0.15  \pm^{   0.08 }_{   0.06 }  $  &  46.7  		\\
\hline
 \end{tabular}
  \label{tab:multivals}
\end{table}

The multipole PDFs derived from the AMIAS analysis were fitted by asymmetric gaussians and numerical results were extracted. 
Table \ref{tab:multivals} lists 
the mean value and  $68\%$ confidence (CL) for 
each of the sensitive multipoles. As sensitive were considered 
all multipole amplitudes with relative uncertainty less than $50\%$. 
The quoted uncertainties of the extracted multipoles include both statistical and systematic errors  and
 contain no model error. The relative uncertainties, 
are also tabulated. The values of the MAID07, SAID-PR15 and BG-2014-02  
models are also shown in the same table for comparison. It is important to highlight the fact that the AMIAS
method exhibits numerical stability even when non-sensitive multipoles are allowed to vary. 
 
\begin{figure} [htp]
\centering
\includegraphics[width=3.5in]{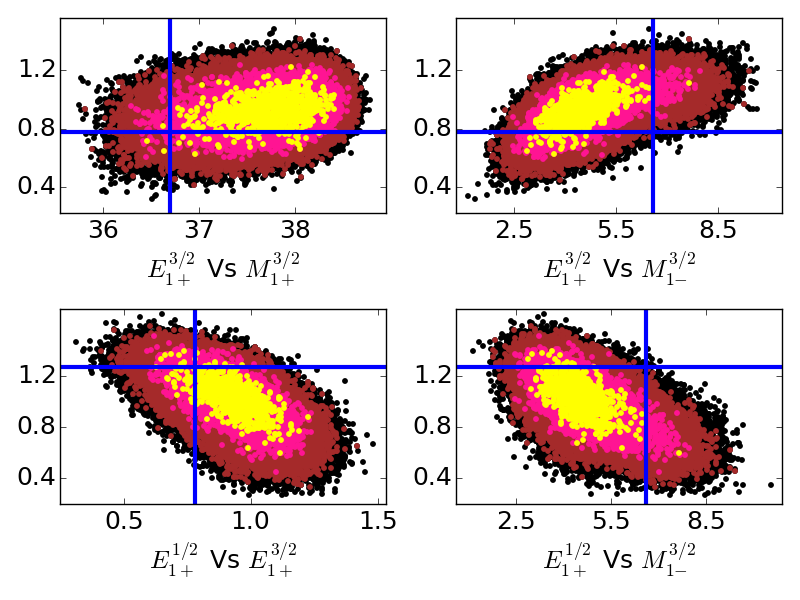}  \hfil 
\vspace*{-6mm}
\caption{Correlation plots between the resonant and  selected background amplitudes. 
The blue lines indicate   the MAID07 model prediction. Color coded as: $\chi^2 \le 1.1\cdot\chi^2_{min}$ (yellow), 
$\chi^2 \le 1.2\cdot\chi^2_{min}$ (pink), $\chi^2 \le 1.3\cdot\chi^2_{min}$ (brown), no $\chi^2$ cut (black).}
\label{fig:mamiepjcorsback}
\end{figure}

\begin{figure} [htp]
 \centering
\includegraphics[width = 3.4in]{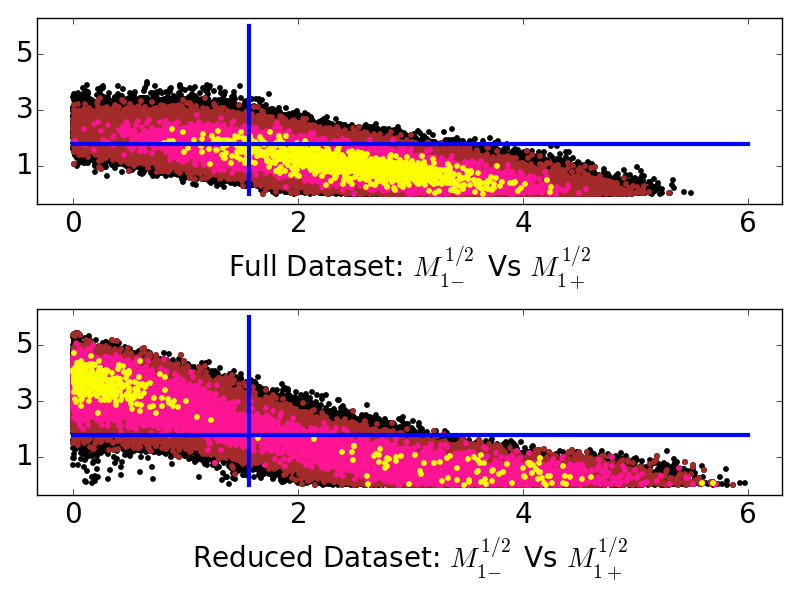} \hfil 
\vspace*{-3mm}
\caption{Correlation plots between the two background amplitudes $M_{1+}^{1/2}$ and $M_{1-}^{1/2}$. 
Top: Extracted from  the 
full database of Table \ref{tab:paperepjdata}. Bottom: Extracted from the reduced dataset.  
The blue cross indicates the MAID07 model prediction. Color coded according to the scheme of fig. \ref{fig:mamiepjcorsback}.}
\label{fig:mamiepjcorsredu}
\end{figure}

 The visualization of the  correlations between any two extracted parameters  is accomplished in a two-dimensional scatter plot in which 
the AMIAS ensemble of solutions are projected on the plane defined by the parameter values and color coded according to the
$\chi^2$ value of each solution. The correlations between the two resonant amplitudes $E_{1+}^{3/2}$ and $M_{1+}^{3/2}$ and the resonant 
and some background amplitudes are shown in Fig. \ref{fig:mamiepjcorsback}. The resonant amplitudes do not exhibit significant correlation 
among them  
whereas there are some mild correlations between the resonant  $E_{1+}^{3/2}$ and  background amplitudes. Some background amplitudes, 
\textit{e.g.} $M_{1+}^{1/2}$ and $M_{1-}^{1/2}$ illustrated in Fig. \ref{fig:mamiepjcorsredu}, exhibit moderate correlations when derived  
from the full dataset 
of Table \ref{tab:paperepjdata} but are  highly correlated when  derived from the ``reduced dataset'' 
which lacks the information of the double polarization $P$ and $G$ observables, thus highlighting the importance of 
double polarization observables.

\subsection{Bands of allowed solutions}
\label{subsec:bands}
The AMIAS ensemble of solutions, can be used to select any subset of  solutions which represent ``reality''   
within a given  CL. This is achieved by  building the histogram 
of the generated  $\chi^2$'s  
forming subsequently from it the corresponding PDF
and  integrating it to the desired CL. In comparison to other approximate methods,  \textit{e.g.} 
the $UP$ parameter method \cite{UP} 
of MINUIT \cite{james1994minuit} the AMIAS method is exact. 
The investigation of the properties of the acceptable solutions can reveal valuable information on the 
 detailed interpretation of the derived results or the potential value of missing measurements. 

\begin{figure*} [htp]
\centering
\includegraphics[width= \textwidth] {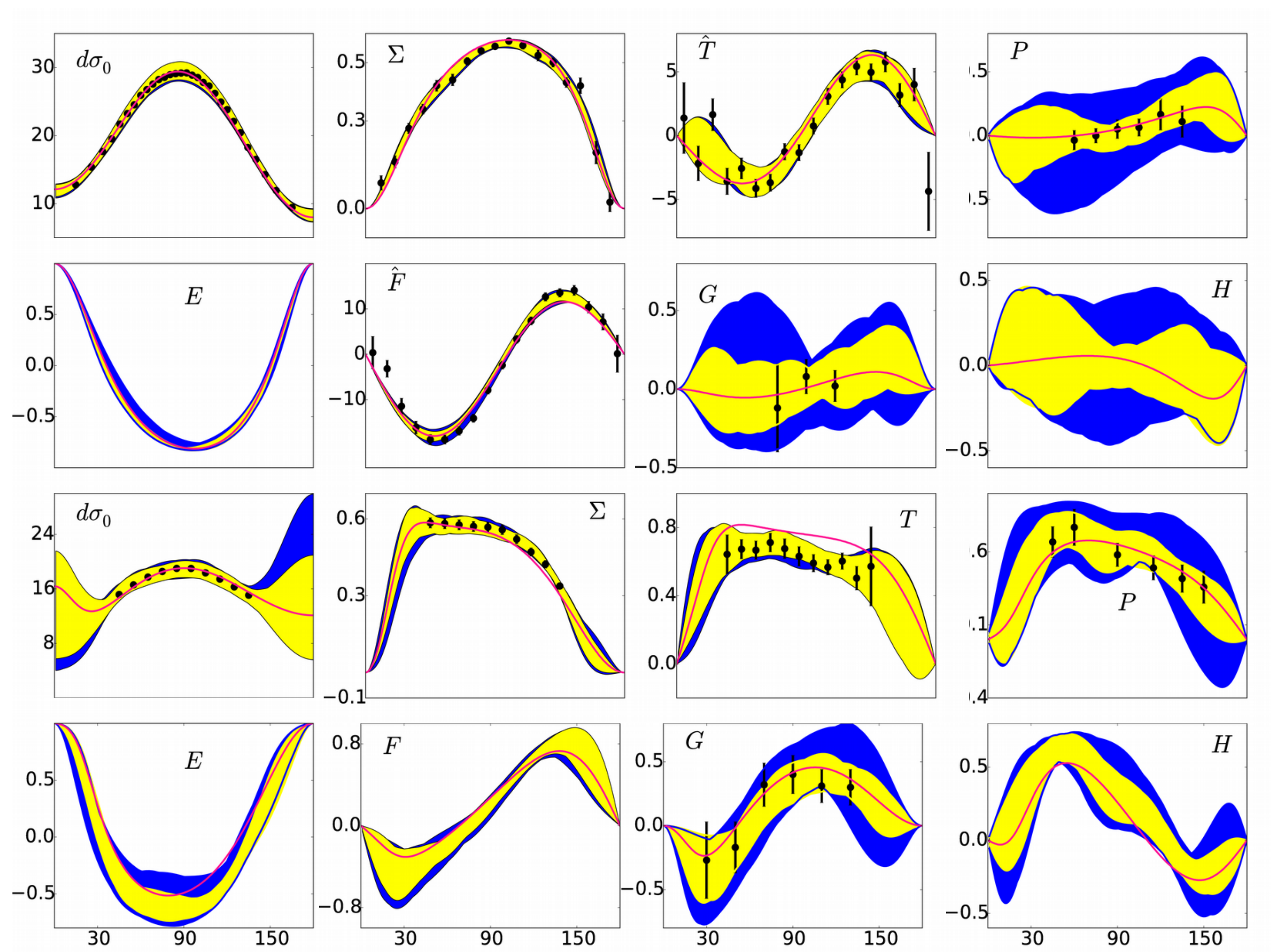} 
\caption{Bands of allowed solutions with a $68\%$ confidence level and a full $\theta_{cm}$ angular coverage 
for the four single and four beam-target pion photoproduction observables. 
The top two rows concern the $\gamma p \to p \pi^0$ channel while the two bottom the $n \pi^+$ channel. Blue bands: the reduced 
dataset is used. Yellow bands: the full dataset is used. The pink curve is the MAID07 model prediction. Cross sections are given 
in units of $\mu  b /sr$.}
\label{fig:paperepj_bands}
\end{figure*}

Using this technique we explore the value of the double polarization observables. 
We selected  the  solutions from the AMIAS ensemble leading to a $68\%$ CL to describing the data. 
 These solutions include values for each of the single 
and double Beam-Target polarization observables which are illustrated as bands in Fig. \ref{fig:paperepj_bands}. 
The yellow more restricted bands correspond to solutions allowed by the full dataset (Table \ref{tab:paperepjdata}) while the blue, 
more relaxed, bands correspond to solutions to a restricted data set  which the double polarization observables $G$ and $P$ have been removed. 
This comparison demonstrates the importance of double polarization observables in restricting the multipole solutions. 
It also indicates the desirability of obtaining  data in the very forward and  backward angles.   
 Such bands, are also particularly  valuable in assessing the value  of observables which need to be measured and the desired accuracy in order 
 to achieve more precise results.

\section{Extracted EMR}
\label{sec:emr}
The derived value of EMR$(\%)$, $-2.5 \pm 0.4_{stat+syst}$,  is free of model error, the first time this has been achieved. 
It is in good agreement 
with earlier reports \cite{beck2000determination,blanpied2001n,ahrens2004helicity}. 
Special care was taken to account for any  systematic errors as described in Section \ref{sec:syst}, for which their effect 
on the derived EMR value was estimated to be $\pm0.1\%$. The magnitude of uncertainty due to systematics was estimated by comparing the 
PDF of EMR when systematic errors were accounted for, and when ignored.

Although a nearly  complete dataset was used,  
which utilizes 
 the most precise measurements to date, 
the derived uncertainty
 is comparable to analyses of older and less precise data.   
This is due to the fact that model dependent analyses in which background multipoles are fixed also freeze the correlations to 
other multipoles \cite{bernstein2007overview} thus reducing the propagated uncertainty. 
This is manifested in the PDFs shown in Figs. \ref{fig:spwaves} and \ref{fig:dwaves} where it is evident that the Model Independent AMIAS 
results (PDFs) are noticeably broader than those resulting from the Model Dependent analysis.
Model independent analysis of the ``benchmark dataset'' \cite{AMIAS_benchmark} has shown 
 that the traditional ansatz \cite{bernstein2007overview} of employing several different models and attributing the spread 
 in the solutions as model error \cite{arndt2001multipole} although  not  precise it adequately captures the magnitude 
 of the effect.

\begin{figure} [h]
\centering
\includegraphics[width=3.4in] {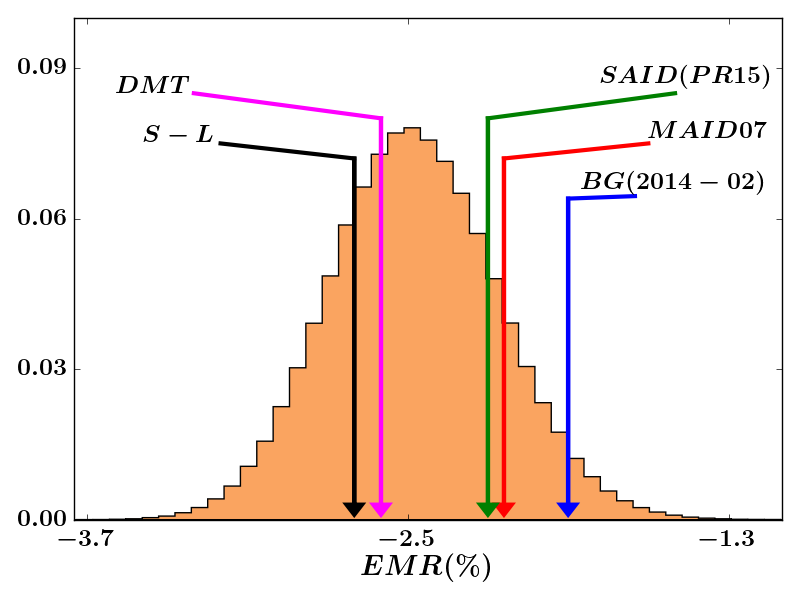} 
\vspace*{-3mm}
\caption{PDF of the derived EMR($\%)$ from the full dataset of Table \ref{tab:paperepjdata}. Both 
statistical and systematic errors were considered. The vertical lines represent model predictions.}
\label{fig:emr_pdf}     
\end{figure}

 In Table \ref{tab:emrvalsepj} we present the EMR values of various analyses and models of the last 20 years and 
make a distinction to the kind of quoted error, statistical, model, and systematic.  Excluding the BRAG result \cite{arndt2001multipole} which 
quotes only a model uncertainty and not a statistical error, all other analyses report a statistical$+$model error combined  
which is comparable or larger than the $\pm 0.4\%$  value determined in this work. Of the listed analyses, only in the works of references 
\cite{beck2000determination,blanpied2001n,ahrens2004helicity} the analyzed data 
allowed for full multipole isospin decomposition and in each of them a different approach was used to fix the background. 
References \cite{beck2000determination} and \cite{ahrens2004helicity}  fix all $\ell \geq 2$ by following the  Born approximation. 
The tabulated EMR value from \cite{beck2000determination}  is from a single energy fit to experimental data while in  \cite{ahrens2004helicity}
EMR was determined using the so-called W-dependent approach which explains the very small statistical errors. Ahrens 
\cite{ahrens2004helicity} 
parameterize  all multipoles which are not resonant by a simple second-order polynomial function with a smooth energy dependence. 
These assumptions contribute to a model error which is estimated by the authors to be $0.3\%$. 
In \cite{blanpied2001n} $(\gamma,\gamma)$  multipoles are also considered, while the $(\pi,\gamma)$ multipole series was 
truncated at $\ell=3$. The general good agreement between the  EMR of this work, the earlier reported values, the latest 
lattice QCD calculation \cite{alexandrou2011nucleon} and 
the model predictions shows that the phenomenological models and the model assumptions 
used up to now are valid. 
Based on this analysis we understand that the observed robustness of the extracted EMR can be 
attributed to the development of successful  phenomenological models but also to the fact that the resonant 
amplitudes exhibit little correlation with the background amplitudes.

\begin{table}  [htp]
\begin{center}
\caption{Selected $EMR(\%)$ values from various experiments, analyses and models reported in the last 20 years.\label{tab:emrvalsepj}}
\scalebox{0.99}{
\begin{tabular}{ll}
        Experiment/Analysis & EMR($\%$)  \\
        \hline
	\textbf{This work}                             & \textbf{-2.5 $\pm$ 0.4 $_{stat+syst}$} $^{(1)}$      \\
	PDG \cite{Patrignani:2016xqp}                  & $-2.5 \pm 0.5$  $^{(2)}$  \\
	Beck '97      \cite{PhysRevLett.78.606}        & $-2.5  \pm 0.2_{stat} \pm 0.2_{model}$    \\
	Blanpied '97 \cite{blanpied1997n}              & $-3.0 \pm 0.3_{stat+syst} \pm 0.2_{model}$                                          \\
	BRAG         \cite{arndt2001multipole}         & $-2.37 \pm 0.27$ $^{(3)}$   \\
	Beck '01 \cite{beck2000determination}          & $-2.5 \pm 0.1_{stat} \pm 0.2_{model}$        \\
	Blanpied '01 \cite{blanpied2001n}              &  $-3.07 \pm 0.26_{stat+syst} \pm 0.24_{model}$                                         \\
	Ahrens '04 \cite{ahrens2004helicity}           &  $-2.74  \pm 0.03_{stat}  \pm 0.3_{model}$ $^{(4)}$ \\
	Kotulla '07 \cite{kotulla2007real}             & $-2.4 \pm 0.16_{stat} \pm 0.24_{model}$   \\
	\hline
	\textbf{Models} & EMR($\%$) \\
	\hline
	Fernandez-Ramirez  \cite{fernandez2006hints}                  & $-3.9 \pm 1.1$ \\
	Pascalutsa - Tjon  \cite{pascalutsa2004pion}                  & $-2.6 \pm 0.6$ \\
	SAID (PR15)  \cite{gwuweb2}                          &  $-2.1$         \\
	Bonn-Gatchina   \cite{bgweb}                &  $-1.9$         \\
	MAID07   \cite{Drechsel:2007if}                &  $-2.2$ 	     \\
	DMT      \cite{kamalov64gamma}                 &  $-2.6$           \\
	Sato-Lee (S-L)   \cite{PhysRevC.63.055201}     &  $-2.7$         \\
	Lattice QCD  \cite{alexandrou2011nucleon} &  $-3.1 \pm 2.1_{stat}$ $^{(5)}$ \\ \hline
     \end{tabular}}
     \begin{tablenotes}
       \small
       \item 1. Model Independent Analysis.
       \item 2. PDG result is an average of several independent reports. 
       \item 3. Quoted uncertainty is purely model and  is the spread \\ of several analyses over the same data.
       \item 4. Quotes BRAG result as model error.
       \item 5. $Q^2 = 0.154(GeV^2), m_{\pi}=297MeV$. 
       \end{tablenotes}
       
 \end{center}
\end{table}

\section{Conclusions}
\label{sec:conclusions}
A dataset of pion photoproduction data which allows for full isospin decomposition was analyzed at a single energy right on top 
of the $\Delta^+(1232)$ resonance. The data contain the most recent and most precise measurements to date.
For the analysis, 
 the AMIAS method was employed 
which allowed the extraction of all  multipole amplitudes to which the data exhibited any sensitivity. 
Our analysis revealed strong 
correlations between background amplitudes and some mild correlations between background amplitudes and the resonant $E_{1+}^{3/2}$ amplitude. 
Since the $E_{1+}^{3/2}$'s determined 
uncertainty dominates the  EMR, capturing all correlations between the parameters was 
an important issue in our analysis which was fully addressed. The $EMR(\%)=-(2.5 \pm 0.4_{stat+syst})$  
reported here is for the first time free of any model error. 
Its good compatibility with phenomenological models and earlier analyses confirms the validity of the model assumptions behind 
the analysis methods used up to now. The model independent results of this work corroborate earlier reports, 
\textit{e.g.} \cite{sparveris2005investigation}, which highlight  
the central role the pion cloud plays in nucleon structure \cite{bernstein2007overview} 
as  a consequence of  the spontaneous  chiral symmetry breaking. 
The derived results, of unprecedented accuracy reconfirm and validate the conjecture of nucleon deformation 
attributing it mostly to pion nucleon dynamical interplay.

\section*{Acknowledgements}
\label{sec:acknowledgements}
The authors would like to thank the $A2$-collaboration for making the recent MAMI measurements available for  analysis. We  are very much indebted to  Reinhard Beck, Michael Ostrick (with support from the Deutsche Forschungsgemeinschaft DFG-CRC1044), Sergey Prakhov and  Yannick Wunderlich 
for enlightening discussions concerning data analysis and experimental and model uncertainties along with Vladimir Pascalutsa, Lothar Tiator and Marc Vanderhaeghen for 
exhaustive discussions on theoretical aspects of nucleon dynamics and nucleon resonance photoexcitation.   
 This work, part of L. Markou Doctoral Dissertation, was supported by the Graduate School of The Cyprus Institute.

\bibliographystyle{unsrt}
\bibliography{bibliography}
\end{document}